\documentclass[prl,twocolumn,showpacs,floatfix,amsfonts]{revtex4}
\usepackage{graphicx,graphics,color,epsfig}
\usepackage{bm}
\usepackage{amsmath}
\usepackage{amssymb}
\usepackage{fancyheadings}
\begin{document}
\preprint{}
\title{
Vortex Core Excitations in Superconductors with Frustrated
Antiferromagnetism}
\author{Jian-Xin Zhu and A. V. Balatsky}
\affiliation{Theoretical Division, Los Alamos National Laboratory,
Los Alamos, New Mexico 87545}
\date{June 10, 2003}
\begin{abstract}
Motivated by recent discovery of cobalt oxide and organic
superconductors, we apply an effective model with strong
antiferromagnetic and superconducting pairing interaction to a
related lattice structure. It is found that the antiferromagnetism
is highly frustrated and a broken-time-reversal-symmetry chiral
$d+id^{\prime}$-wave pairing  state prevails. In the mixed state,
we have solved the local electronic structure near the vortex core
and found no local induction of antiferromagnetism. This result is
in striking contrast to the case of copper oxide superconductors.
The calculated local density of states indicates the existence of
low-lying quasiparticle bound states inside the vortex core, due
to a fully gapped chiral pairing state. The prediction can be
directly tested by scanning tunneling microscopy.
\end{abstract}
\pacs{74.70.-b, 74.20.Rp, 74.20.-z, 74.25.Jb} \maketitle Recently,
superconductivity in the oxyhydrate Na$_{0.35}$CoO$_{2}$ $\cdot$
1.3H$_{2}$O below $\sim 5K$ was discovered by Takata {\em et
al.}~\cite{Takada03} and confirmed immediately by several
groups~\cite{Foo03,Lorentz03,Rivadulla03,Schaak03}. Interesting
features of this material include: (i) It is believed that the
superconductivity comes from CoO$_{2}$ layers, similar to that in
copper-oxide cuprates. (ii) Co$^{4+}$ atoms have
spin-$\frac{1}{2}$ but form a triangular lattice, which frustrates
the antiferromagnetism (AF) and thus is a promising candidate for
the occurrence of spin-liquid phases~\cite{Anderson73}. There are
also organic superconductors like the $\kappa$-(BEDT-TTF)$_{2}$X
materials which have a lattice structure very similar to the
triangular lattice~\cite{McKenzie97,Kontani03}. (iii)
Theoretically, the analysis based on the resonant valence bond
theory in the framework of the $t$-$J$
model~\cite{Baskaran03,Kumar03,Wang03,Ogata03} or on the
renormalization group theory within the framework of $t$-$U$-$J$
model~\cite{Honerkamp03} indicates a wide window of
broken-time-reversal-symmetry (BTRS) $d$+$id^{\prime}$-wave
pairing state in the phase diagram. Other theoretical
groups~\cite{Tanaka03,Singh03,Ni03} proposed a $p_{x}+ip_{y}$-wave
pairing state mediated by ferromagnetic fluctuations. Since the
ferromagnetism is insensitive to the detailed lattice structure,
no frustration effect is expected on a triangular lattice.
\begin{figure}
\centerline{\psfig{figure=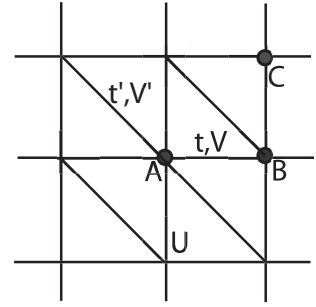,height=4cm,width=4cm,angle=0}}
\caption[*]{Lattice structure of a superconductor with frustrated
antiferromagnetism. It has a nearest-neighbor hopping $t$ and
pairing interaction $V$ on the bonds forming the two-dimensional
lattice, a next-nearest-neighbor hopping integral $t^{\prime}$ and
$V^{\prime}$ along {\em one} diagonal of each plaquette, and an
on-site Hubbard interaction $U$ on each site. The sites $A$, $B$,
and $C$ labelled by filled circles are the vortex core center and
its neighbors when the superconductor is in the mixed state. }
\label{FIG:Lattice}
\end{figure}
At this stage, we are not at a position to resolve the pairing
state issue. Motivated by recent observation of a superconducting
phase diagram of Na$_{x}$CoO$_{2}\cdot 1.3$H$_{2}$O similar to
that of the cuprate superconductors~\cite{Schaak03}, we consider
in this Letter a spin-singlet pairing and study the nature of
low-lying excitations around a vortex in these new superconductors
with frustrated antiferromagnetism. The results can be directly
tested by further experiments such as scanning tunneling
microscopy (STM), which likely will  be carried out soon.

In conventional $s$-wave superconductors such as NbSe$_{2}$, the
observed quasiparticle tunneling spectrum at the vortex core by
Hess {\em et al.}~\cite{Hess89} can be explained successfully in
terms of the low-lying quasiparticle bound states as shown by
Caroli, de Gennes, and Matricon~\cite{Caroli64}. In copper-oxide
cuprates, the condensate has a $d$-wave pairing symmetry.
Theoretical study based on $d$-wave BCS model
suggested~\cite{Wang95} that, due to the existence of nodal
quasiparticles,  the local density of states (LDOS) at the
$d$-wave vortex core exhibits a single broad peak at zero energy.
However, the STM-measured local differential tunneling conductance
at the vortex core center only exhibits a subgap double-peak
structure~\cite{Maggio95,Pan00} or even no clear peak structure
within the superconducting gap~\cite{Renner98}. The discrepancy
between the earlier theoretical prediction and the experimental
observation stimulated various explanations. Recent intensive
experimental~\cite{Lake01,Hoffman02,Mitrovic01,Khaykovich03,Kakuyanagi03}
and
theoretical~\cite{Demler01,Zhu01,Chen02,Franz02,Takigawa03,Andersen03}
studies seem to converge on an explanation in terms of the
field-induced AF around the vortex core. When the AF is frustrated
on a triangular lattice~\cite{Tsai01,Vojta99}, one would expect a
different nature of electronic excitations near the vortex.
Previously, we have applied an effective microscopic mean-field
model with competing AF and superconducting interactions to a
square lattice, as relevant to the copper-oxide superconductors.
This model generates very rich physics including the commensurate
AF spin density wave ordering at undoped systems, stripes at low
doping, as well as the superconducting states at optimal and
overdoped regimes~\cite{Martin00}. Especially, within this model,
it was found~\cite{Zhu01} that the AF ordering is induced around
the vortex core, which explains several experimental observations
on cuprates. Here we extend this effective model to a lattice as
shown in Fig.~\ref{FIG:Lattice}, which interpolates between the
square and triangular lattices. Analysis of this paper is also
directly applicable to organic conductors that have such a lattice
structure.

The model consists of an on-site repulsion and off-site
attraction. The former is solely responsible for the
antiferromagnetism while the latter causes the superconductivity.
The mean-field Hamiltonian is written as:
\begin{eqnarray}
H&=&-\sum_{ij,\sigma} t_{ij}e^{i \varphi_{ij}}
c_{i\sigma}^{\dagger}c_{j\sigma} +\sum_{i,\sigma} (U
n_{i,\bar{\sigma}}+\epsilon_{i}-\mu)
c_{i\sigma}^{\dagger}c_{i\sigma} \nonumber \\
&&+\sum_{ij} (\Delta_{ij} c_{i\uparrow}^{\dagger}
c_{j\downarrow}^{\dagger} +\Delta_{ij}^{*} c_{j\downarrow} c_{
i\uparrow} ) \;. \label{EQ:MFA}
\end{eqnarray}
Here  $c_{i\sigma}$ annihilates an electron of spin $\sigma$ at
the $i$th site. The hopping integrals are respectively $t_{ij}=t$
on the bonds forming the two-dimensional (2D) lattice and
$t^{\prime}$ along {\em one} diagonal of each plaquette. The
on-site repulsion is $U$ on each site. The quantities
$n_{i\sigma}=\langle c_{i\sigma}^{\dagger} c_{i\sigma}\rangle$,
$\epsilon_i$, and $\mu$, are the electron density with spin
$\sigma$, the single site potential describing the scattering from
impurities, and the chemical potential. The spin-singlet order
parameter $\Delta_{ij}=\frac{V_{ij}}{2}\langle c_{i\uparrow}
c_{j\downarrow} -c_{i\downarrow}c_{j\uparrow}\rangle$ comes from
the pairing interactions on the bond ($V_{ij}=V$) and along {\em
one} diagonal of each plaquette ($V_{ij}=V^{\prime}$). The case of
$t^{\prime}=0$ and $V^{\prime}=0$ correspond to the model on a
square lattice. With the application of an external magnetic field
$\mathbf{H}$, the Peierls phase factor is given by the integral
$\varphi_{ij}=\frac{\pi}{\Phi_{0}} \int_{{\bf r}_{j}}^{{\bf
r}_{i}} \mathbf{A}({\bf r})\cdot d{\bf r}$, where  the
superconducting flux quantum $\Phi_0=hc/2e$ and
 the vector potential $\mathbf{A}=\frac{1}{2}
 \mathbf{H}\times \mathbf{r}$ in the symmetric gauge. The
 Hamiltonian~(\ref{EQ:MFA}) can also be derived from the $t$-$U$-$J$
 model~\cite{Honerkamp03,Daul00}.
We diagonalize Eq.~(\ref{EQ:MFA}) by solving self-consistently the
Bogoliubov-de Gennes equation:
\begin{equation}
\sum_{j} \left(
\begin{array}{cc}
{\cal H}_{ij,\sigma} & \Delta_{ij}  \\
\Delta_{ij}^{*} & -{\cal H}_{ij,\bar{\sigma}}^{*}
\end{array}
\right) \left(
\begin{array}{c}
u_{j\sigma}^{n} \\ v_{j\bar{\sigma}}^{n}
\end{array}
\right) =E_{n} \left(
\begin{array}{c}
u_{i\sigma}^{n} \\ v_{i\bar{\sigma}}^{n}
\end{array}
\right)  \;, \label{EQ:BdG}
\end{equation}
subject to the self-consistency conditions for the electron
density and the superconducting order parameter:
$n_{i\uparrow}=\sum_{n} \vert u_{i\uparrow}^{n}\vert^{2} f(E_n)$
and $n_{i\downarrow}= \sum_{n}\vert
v_{i\downarrow}^{n}\vert^{2}[1-f(E_n)]$, and
$\Delta_{ij}=\frac{V_{ij}}{4}\sum_{n} (u_{i\uparrow}^{n}v_{
j\downarrow}^{n*} +v_{i\downarrow}^{n*}u_{j\uparrow}^{n} ) \tanh
\left( \frac{E_{n}}{2k_{B}T}\right)$. Here the quasiparticle
wavefunction, corresponding to the eigenvalue $E_n$, consists of
the component $u_{i\sigma}^{n}$ for an electron of spin $\sigma$
and the component $v_{i\bar{\sigma}}^{n}$ for a hole of opposite
spin $\bar{\sigma}$. The single particle Hamiltonian reads ${\cal
H}_{ij,\sigma}=-t_{ij} e^{i\varphi_{ij}} +(Un_{i\bar{\sigma}}
+\epsilon_{i}-\mu)\delta_{ij}$. The Fermi distribution function is
$f(E)=1/[e^{E/k_{B}T}+1]$.  Hereafter we measure the length in
units of the lattice constant $a_0$ and the energy in units of $t
(>0)$. As relevant to the superconductivity in the cobalt oxides,
we report results below for the case with $t^{\prime}=t$ and
$V^{\prime}=V$. As a model calculation, we choose $U=4$ and
various values of $V$. We find that the antiferromagnetism is
highly frustrated even when $V=0$ and the filling factor is one
electron per site. Since the experiments on cobalt oxides were
performed in the optimal or slightly overdoped
regime~\cite{Takada03,Schaak03}, we choose the filling factor
$n_f=\sum_{i,\sigma} n_{i\sigma} /N_x N_y=0.65$, where $N_x,N_y$
are the linear dimensions of the unit cell under consideration.
The chosen band filling factor corresponds to an electron doping
$x=0.35$~\cite{Honerkamp03}. We use an exact diagonalization
method to solve the BdG equation~(\ref{EQ:BdG}) self-consistently.
In zero field, the solution is found to be uniform: There exits no
AF spin density wave (SDW) and the superconducting bond order
parameters exhibit the relation that $\Delta_{x}=\vert
\Delta_{0}\vert e^{-i\theta}$, $\Delta_{y}= \vert \Delta_{0} \vert
e^{i\theta}$, $\Delta_{xy}=\vert \Delta_{0} \vert $ with
$\theta=2\pi/3$, consistent with the results based on the $t$-$J$
model (see, e.g., \cite{Ogata03}). We find typically $\vert
\Delta_{0}\vert =0.07$ and 0.15 for $V=2.5$ and $3.0$,
respectively. This complex order parameter forms a
broken-time-reversal-symmetry chiral pairing state. We then use
this zero-field solution as the initial condition for iteration
for the vortex problem.
\begin{figure}
\centerline{\psfig{figure=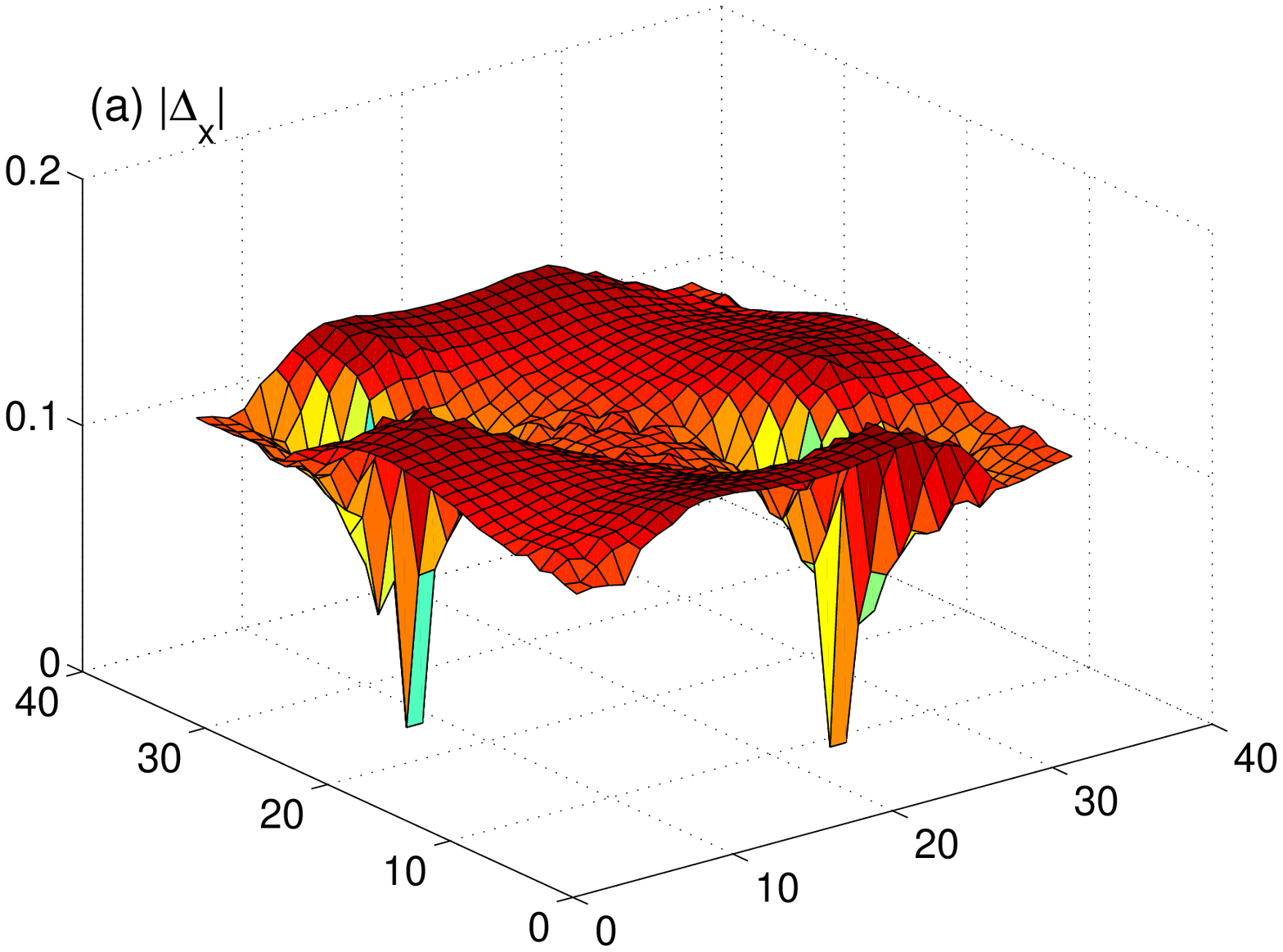,height=4.0cm,width=5.0cm,angle=0}}
\centerline{\psfig{figure=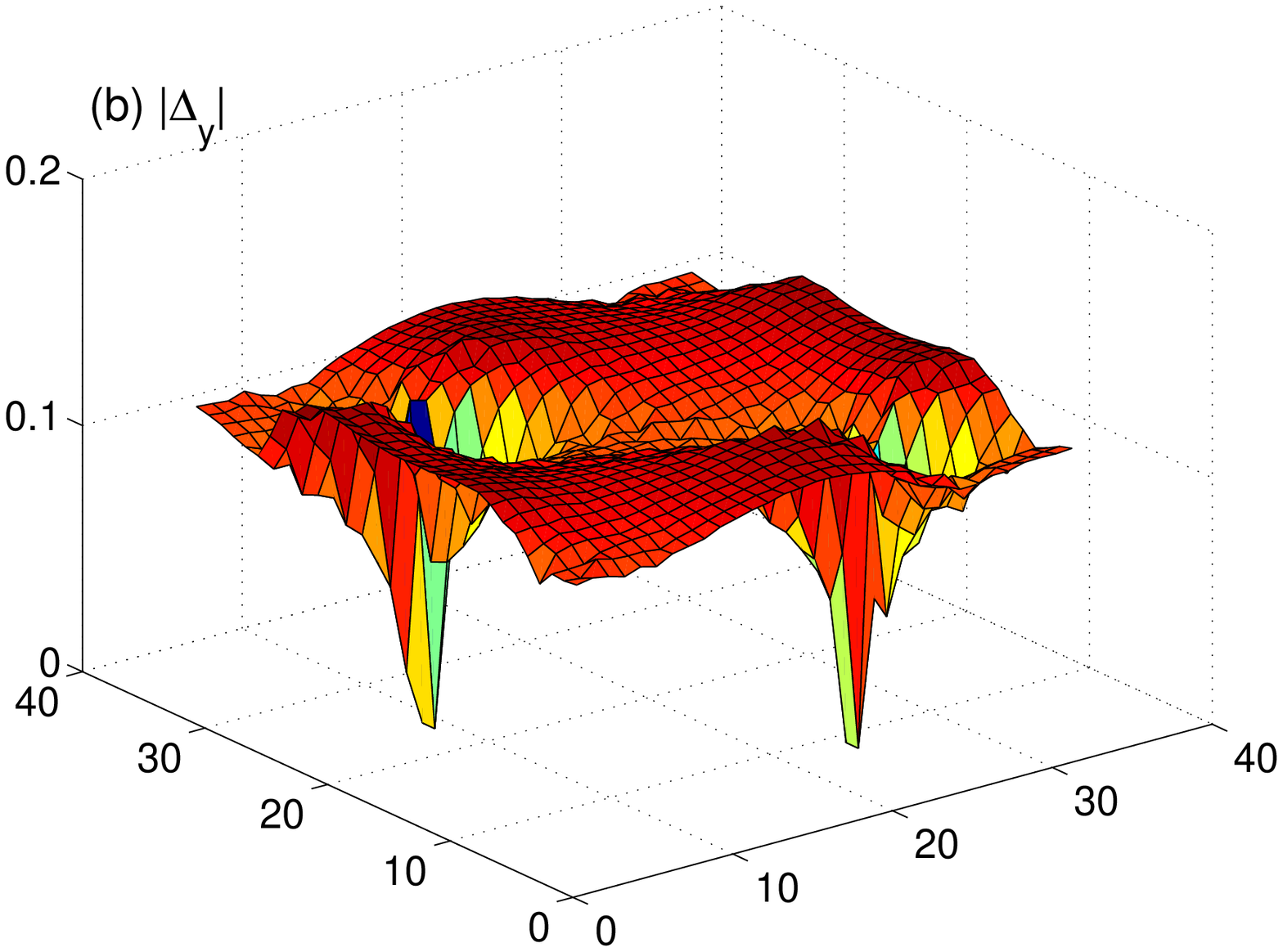,height=4.0cm,width=5.0cm,angle=0}}
\centerline{\psfig{figure=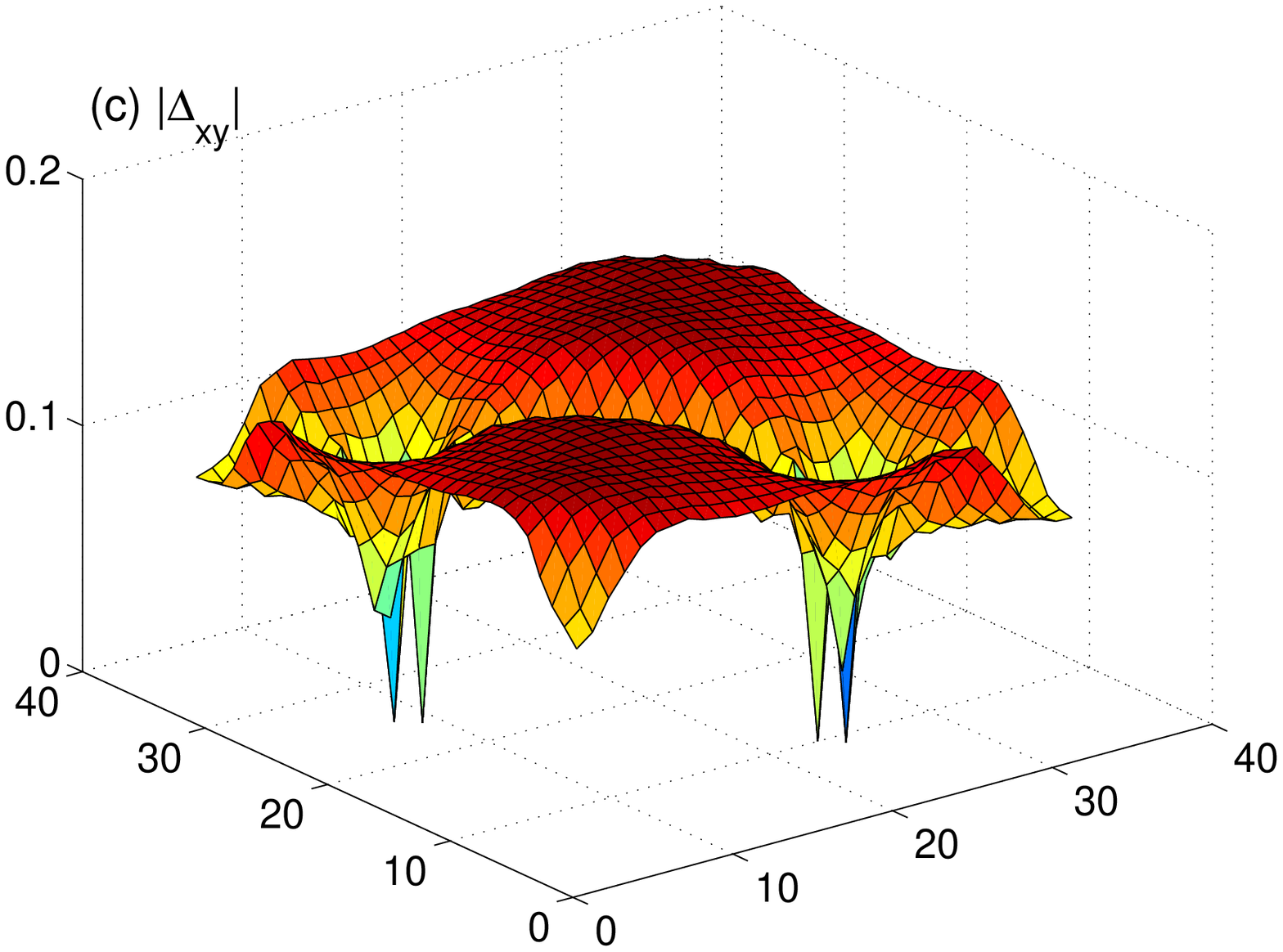,height=4.0cm,width=5.0cm,angle=0}}
\centerline{\psfig{figure=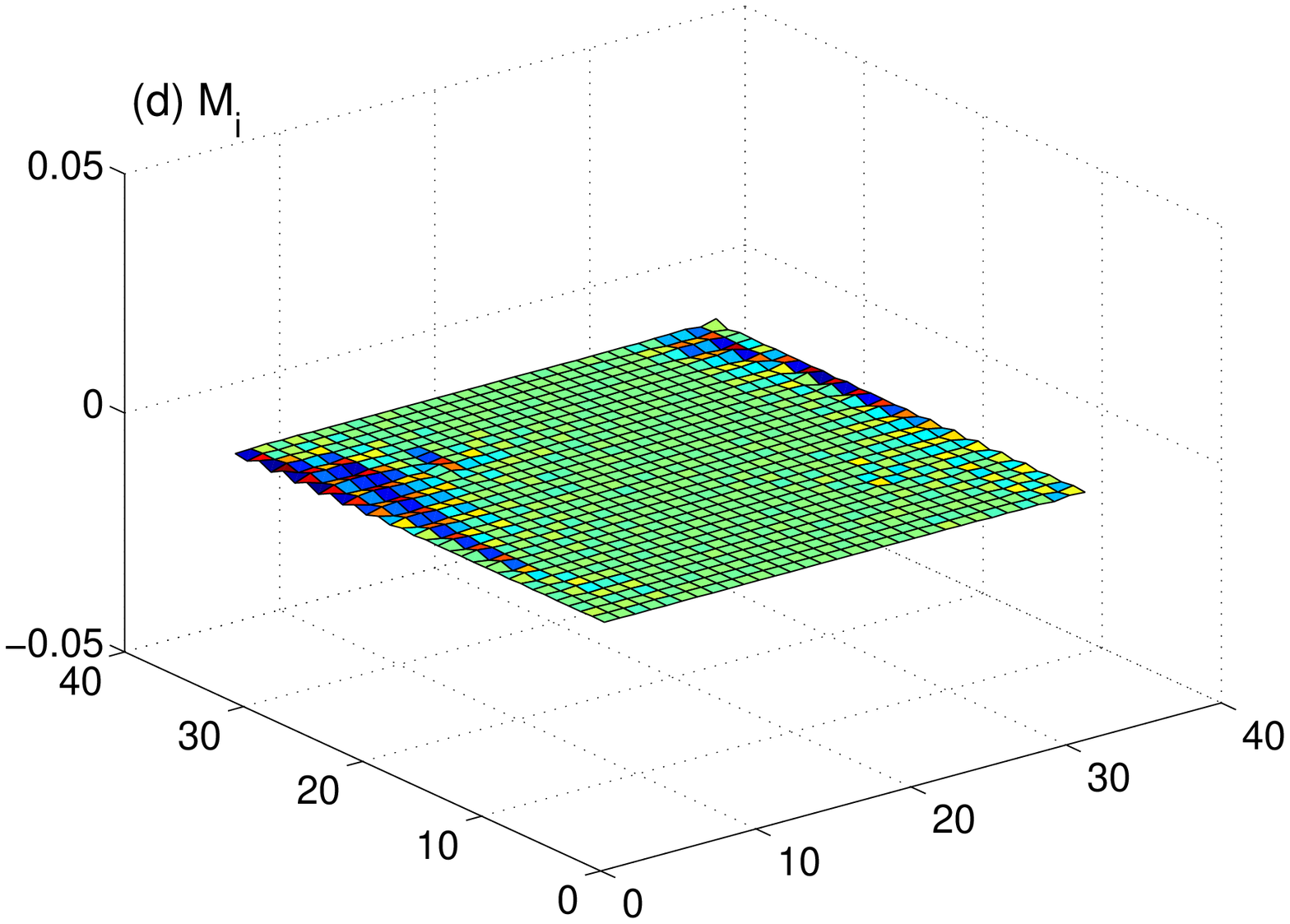,height=4.0cm,width=5.0cm,angle=0}}
\centerline{\psfig{figure=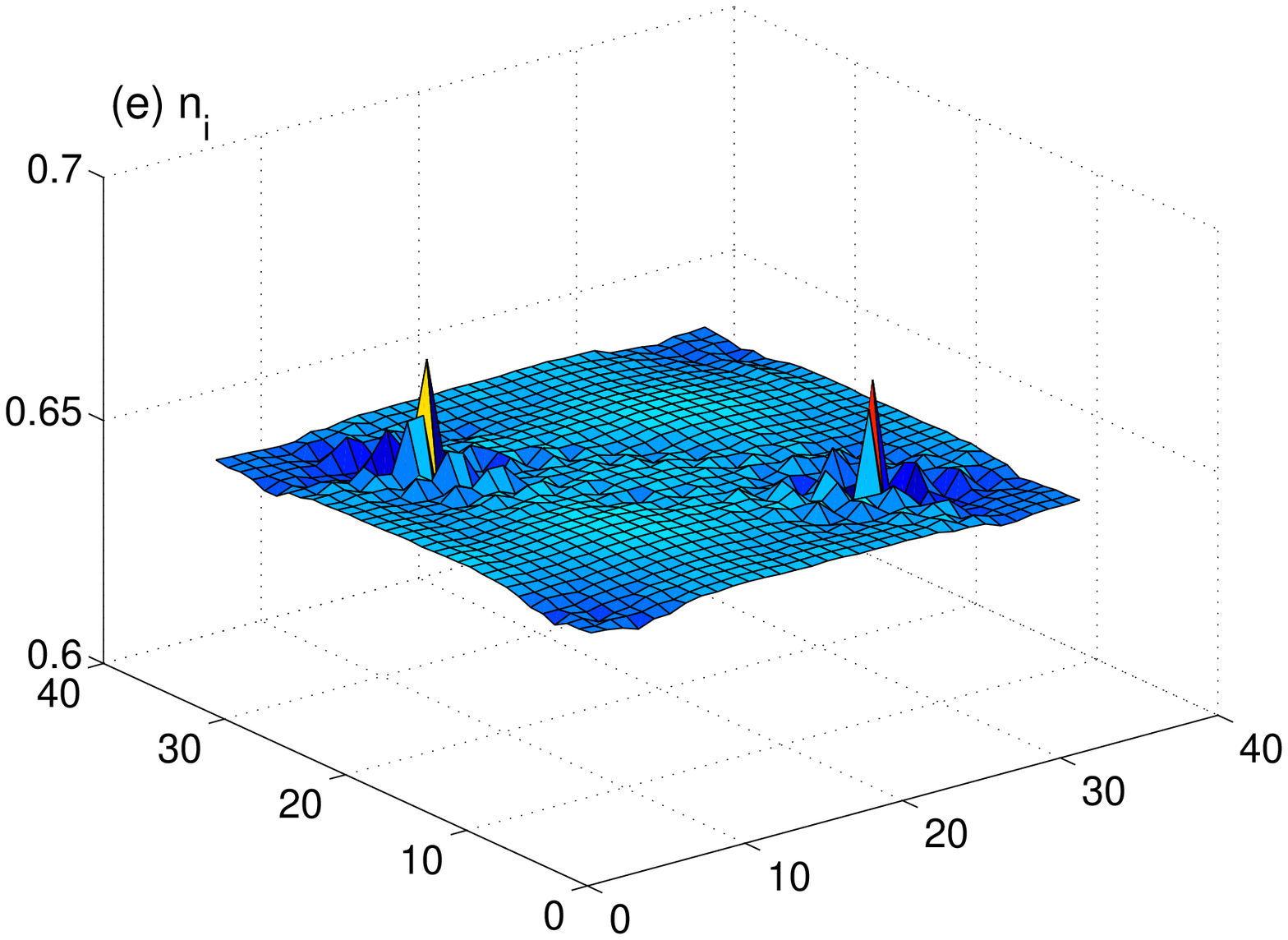,height=4.0cm,width=5.0cm,angle=0}}
\caption[*]{The three-dimensional display of the amplitude
distribution of (a-c) the three superconducting bond order
parameters, (d) the staggered magnetization $M_{i}$, and (e) the
electron density $n_{i}=\sum_{\sigma}n_{i\sigma}$ in a magnetic
unit cell containing two vortices. The size of the unit cell is
$32\times 32$. Parameter values: $U=4$, $V=3$, $n_f=0.65$, and
$T=0$. } \label{FIG:VORTEX}
\end{figure}

When an external magnetic field is applied perpendicular to the 2D
Co-Co plane, $\mathbf{H}=H\hat{\mathbf{z}}$ ($H_{c1}\ll H\ll
H_{c2}$), an Abrikosov vortex state is formed. Recent experiments
suggest~\cite{Sakurai03,Cao03} that  the cobalt-oxide
superconductor is in the extreme type-II limit (the
Ginzburg-Landau parameter $\kappa>100$). Therefore, the screening
effect from the supercurrent is negligible. By taking the strength
of magnetic field, $H=2\Phi_0/N_x N_y$, such that the total flux
enclosed is twice $\Phi_0$, the BdG equation~(\ref{EQ:BdG}) is
solved with the aid of the magnetic Bloch theorem~\cite{Zhu01}:
\begin{equation}
\left(
\begin{array}{c}
u_{{\bf k},\sigma}({\cal T}_{mn}\mathbf{r}) \\
v_{{\bf k},\sigma}({\cal T}_{mn}\mathbf{r})
\end{array}
\right)
 = e^{i{\bf k}\cdot {\bf R}}
\left(
\begin{array}{c}
e^{i\chi({\bf r},{\bf R})/2} u_{{\bf k},\sigma}(\mathbf{r}) \\
e^{-i\chi({\bf r},{\bf R})/2}v_{{\bf k},\sigma}(\mathbf{r})
\end{array}
\right) \;.
\end{equation}
Here $\mathbf{r}$ is the position vector defined within a given
unit cell, the vector ${\bf R}=m N_x \hat{\bf e}_{x} +n N_y
\hat{\bf e}_{y}$, ${\bf k}=\frac{2\pi l_x}{M_x N_x}\hat{\bf e}_{x}
+ \frac{2\pi l_y}{M_y N_y}\hat{\bf e}_{y}$ with
$l_{x,y}=0,1,\dots,M_{x,y}-1$ are the wavevectors defined in the
first Brillouin zone of the vortex lattice, $M_x N_x$ and $M_y
N_y$ are  the linear dimension of the whole system, and the phase
$\chi({\bf r},{\bf R})=\frac{2\pi}{\Phi_0}{\bf A}({\bf R}) \cdot
{\bf r} +2mn\pi$. For the calculation, we consider a single
magnetic unit cell of size $N_x\times N_y=32\times 32$. Typical
results on the nature of the vortex core is displayed in
Fig.~\ref{FIG:VORTEX} with $V=3$. It shows that each unit cell
accommodates two superconducting vortices each carrying a flux
quantum $\Phi_0$, which conforms to the above prescription for the
magnetic field strength. The two vortices sit evenly on the
diagonal of the unit cell, indicative of a diagonal-oriented
square vortex. As shown in Fig.~\ref{FIG:VORTEX}(a)-(c), each
component of the superconducting order parameter vanishes at the
vortex core center and starts to increase to its bulk value, as
one moves away from the core. Fig.~\ref{FIG:VORTEX}(d) displays
the spatial distribution of the staggered magnetization of the
local SDW order as defined by $M_{i}=(-1)^{i}
(n_{i\uparrow}-n_{i\downarrow})$. It is vanishingly small [A weak
variation in the figure is due to the numerical accuracy]. The
electron density $n_{i}=\sum_{\sigma} n_{i\sigma}$ exhibits a
Friedel oscillation around the vortex core
(Fig.~\ref{FIG:VORTEX}(e)), in contrast to a monotonic decrease of
the electron density when a BCS $d$-wave vortex is approached.
This is a direct manifestation of the vortex bound states inside
the core. Finally, the vortex core shows a twofold symmetry,
consistent with the underlying lattice structure under
consideration. It is natural to expect that the vortex core on a
triangular lattice would exhibit a sixfold symmetry. Most
importantly, we find no local AF ordering around the vortex core
in a superconductor with a strong AF interaction, and therefore
the vortex core is predicted to be a conventional one. Our results
here are fundamentally different from the case in cuprates, where
a local AF ordering is nucleated around the vortex
core~\cite{Lake01,Hoffman02,Mitrovic01,Khaykovich03,Kakuyanagi03,Demler01,Zhu01,Chen02,Franz02,Takigawa03,Andersen03}.
\begin{figure}
\centerline{\psfig{figure=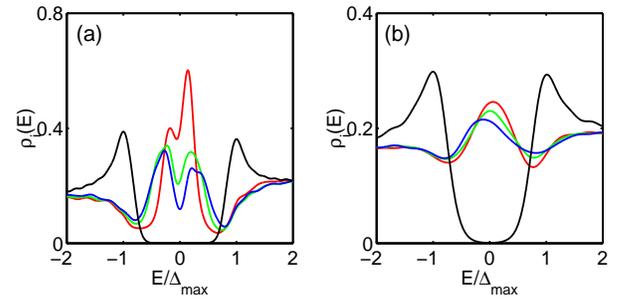,height=4.0cm,width=8.0cm,angle=0}}
\caption[*]{The local density of states as a function of energy at
three selected sites A (red line), B (green line), and C (blue
line) for (a) $V=3$ and (b) $V=2.5$. The relative positions of the
selected sites are indicated on Fig.~\ref{FIG:Lattice}. Also shown
is the density of states for the zero-field uniform case. The
energy is scaled to the maximum superconducting gap
$\Delta_{max}$. The other parameter values are the same as in
Fig.~\ref{FIG:VORTEX} except for $T=0.02$.
 } \label{FIG:LDOS}
\end{figure}

The LDOS is defined by
\begin{eqnarray}
\rho_{i}(E)&=&-\frac{1}{M_x M_y} \sum_{\mathbf{k},n,\sigma} [\vert
u_{i\uparrow}^{n,\mathbf{k}} \vert^{2}
f^{\prime}(E^{n,\mathbf{k}}-E) \nonumber \\
&&+\vert v_{i\downarrow}^{n,\mathbf{k}}\vert^{2}
f^{\prime}(E^{n,\mathbf{k}}+E)]\;,
\end{eqnarray}
where $f^{\prime}(E)$ is the derivative of the Fermi distribution
function. The $\rho_{i}(E)$ is proportional to the local
differential tunneling conductance which could be measured by STM
experiments. 
In Fig.~\ref{FIG:LDOS} we plot the LDOS as a function of energy at
the three selected sites around the vortex core center, which are
labelled in Fig.~\ref{FIG:Lattice}  for various  values of $V$.
For comparison, we have also displayed the density of states for
the zero-field uniform case, which is peaked at the maximum
superconducting energy gap $\Delta_{max}$. The coherent peaks at
the gap edge are smeared out in the LDOS at the vortex core.
Instead there are sharp subgap resonance peaks in the LDOS. Due to
the energy level quantization of the bound states, the small gap
around the Fermi energy follows approximately the relation,
$\Delta_{1}\sim \Delta_{max}^{2}/E_{F}$, similar to the $s$-wave
case. For a strong pairing interaction, which leads to a large
magnitude of the superconducting order parameter, the energy level
spacing is wide (Fig.~\ref{FIG:LDOS}(a)). For a weak pairing
interaction, the level spacing is so small that only a thermally
broadened peak is observed. This result is dramatically different
from that in a BCS $d$-wave superconductor. In the latter, due to
the existence of nodal quasiparticles, there always appears a
broadened single peak around the Fermi energy regardless of how
short the superconducting coherence length is. We propose to use
the STM to investigate the core structure in these materials. For
cobalt oxides, the superconducting energy would be very small,
since $T_{c}\sim  5K$ at slightly overdoped regime. It is expected
that a broadened single peak should be observed. It is a
possibility that cobalt oxide superconductors have a spin-triplet
pairing symmetry mediated by ferromagnetic fluctuations. One would
expect therefore a competition between ferromagnetism and
superconductivity in this case and vortex core excitations
analogous to the ones in copper oxide cuprates may occur. This
question is reserved for an interesting future study.

\begin{figure}
\centerline{\psfig{figure=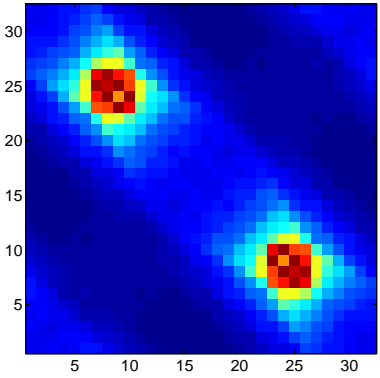,height=5.0cm,width=5.0cm,angle=0}}
\caption[*]{The spatial distribution of the LDOS at
$E=-0.27\Delta_{max}$, at which the subgap resonance peak shows up
in the LDOS at site $C$ for $V=3.0$. The other parameter values
are the same as Fig.~\ref{FIG:VORTEX} except for $T=0.02$.}
\label{FIG:IMAGE}
\end{figure}

We now look into the nature of the vortex states.  In
Fig.~\ref{FIG:IMAGE}, we display the spatial distribution of the
local density of states at a resonant energy level. It shows no
long-range tails, in striking contrast to the case of $d$-wave
superconductors, where tails runs along the gap nodal directions.
Instead, the LDOS at the core level is trapped in an area
characterized by the superconducting coherence length, indicating
that the core states are completely localized.

In conclusion, we have studied the vortex core excitations in
superconductors with frustrated antiferromagnetism. This
investigation is relevant to recently discovered cobalt oxide and
organic  superconductors. We find no local induction of
antiferromagnetism around the vortex core. Low-lying quasiparticle
bound states are found inside the vortex core due to a fully
gapped chiral pairing state. These core states can be directly
measured by STM experiments regardless of the magnitude of the
level spacing between them. To our knowledge, this is the first
theoretical investigation of the nature of vortex core states in
superconductors with frustrated antiferromagnetism.

We thank Y. Bang, M. Graf, Y. N. Joglekar, Z. Nussinov, and S. B.
Shastry for useful discussions. This work was supported by the US
Department of Energy.

\end{document}